\documentclass{PoS}

\newcommand\cF{{c}_F}
\newcommand\cS{{c}_\Sigma}
\newcommand\cM{{c}_M}

\usepackage{amsmath}

\title{$\sigma$-resonance and convergence of chiral perturbation theory}

\ShortTitle{$\sigma$-resonance and convergence of CHPT}

\author{D.~J.~Cecile}

\author{\speaker{Shailesh Chandrasekharan}\thanks{This work was supported in part by the Department of Energy grant DE-FG02-05ER41368.} \\ \\
         Department of Physics, Box 90305, 
         Duke University, Durham, North Carolina 27708.\\
         E-mail: \email{sch@phy.duke.edu}}

\abstract{The dimensionless parameter $\xi' = M^2/(16 \pi^2 F^2)$, where $F$ is the pion decay constant in the chiral limit and $M$ is the pion mass at leading order in the quark mass, is expected to control the convergence of chiral perturbation theory applicable to QCD. Here we demonstrate that a strongly coupled lattice gauge theory model with the same symmetries as two-flavor QCD but with a much lighter $\sigma$-resonance is different. Our model allows us to study efficiently the convergence of chiral perturbation theory as a function of $\xi$. We first confirm that the leading low energy constants appearing in the chiral Lagrangian are the same when calculated from the $\epsilon$-regime and the $p$-regime. However, $\xi' \lesssim 0.002$ is necessary before 1-loop chiral perturbation theory predicts the data within 1\%. However, for $\xi' > 0.0035$ the data begin to deviate qualitatively from 1-loop chiral perturbation theory predictions. We argue that this qualitative change is due to the presence of a light $\sigma$-resonance in our model. Our findings may be useful for lattice QCD studies.}

\FullConference{The XXVI International Symposium on Lattice Field Theory\\
		 July 14-19 2008\\
		 Williamsburg, Virginia, USA}

\begin{document}

\section{Introduction}

Chiral perturbation theory is an effective field theory that describes the low momentum physics of the pions. According to this theory physical quantities can be expressed as a power series in the dimensionless parameter $\xi' = M^2/(16 \pi^2 F^2)$ where $M^2 = m \Sigma/F^2$ is the square of the pion mass to the leading order in the quark mass $m$, $F$ is the pion decay constant and $\Sigma$ is the chiral condensate both evaluated in the chiral limit. The unknown coefficients of the power series encode the dynamics of QCD and are referred to as the low energy constants. One of the important topics of research today is to match lattice QCD data to the chiral expansion and compute these low energy constants from first principles \cite{Necco:2007pr,Dimopoulos:2007qy}. In order to match data with the chiral expansion reliably it is important to find the range in $\xi$ where the predictions will be valid \cite{Bernard:2002yk,Giusti:2007hk}. The region of validity is governed by the properties of resonances in the theory. Can the $\sigma$-resonance affect the convergence of the chiral expansion?

The $\sigma$-resonance arises in $\pi-\pi$ scattering in a channel with vacuum quantum numbers. In the physical world sigma is a broad resonance and is usually ignored in the analysis. Recently, it was estimated that $M_\sigma \simeq 440$MeV and $\Gamma_\sigma \simeq 544$MeV \cite{Caprini:2005zr}. On the other hand, in lattice QCD, as the pion masses increase, the properties of the resonance will clearly change. Indeed recent studies do find a strong dependence on the quark mass \cite{Hanhart:2008mx}. It is interesting to ask if this dependence can affect the chiral expansion. This question is non-perturbative and difficult to answer within lattice QCD currently. On the other hand it may be answerable in models with the same symmetries as QCD and such studies may help shed some light on the subject.

Here we study a QCD-like lattice field theory model which has the same symmetries as two-flavor QCD. Our model also contains a parameter which we tune so that it contains a light sigma resonance in addition to pions. We then find evidence that indeed chiral perturbation theory breaks down when $M_\pi \gtrsim M_\sigma/3$. For more details we refer to the published versions of this work in \cite{Cecile:2007dv,Cecile:2008kp}.

\section{Model and Observables}

Our model involves two flavors of staggered fermions interacting strongly with abelian gauge fields. The action of the model is given by 
\begin{equation}
S = - \sum_{x}\sum_{\mu=1}^{5}
\eta_{\mu,x}\bigg[\mathrm{e}^{i\phi_{\mu,x}}{\overline\psi}_x
{\psi}_{x+\hat\mu}
-\mathrm{e}^{-i\phi_{\mu,x}}{\overline\psi}_{x+\hat\mu}{\psi}_x\bigg] 
- \sum_x \bigg[m{\overline\psi}_x{\psi}_x
+\frac{\tilde c}{2}\bigg({\overline\psi}_x{\psi}_x \bigg)^2\bigg],
\label{eq1}
\end{equation}
where $x$ denotes a lattice site on a $4+1$ dimensional hyper-cubic lattice $L_t \times L^4$.  Here $L^4$ is the usual Euclidean space-time box while $L_t$ represents a fictitious temperature direction whose role will be discussed below. The two component Grassmann fields, $\overline\psi_x$ and $\psi_x$, represent the two quark $(u,d)$ flavors of mass $m$, and $\phi_{\mu,x}$ is the compact $U(1)$ gauge field through which the quarks interact. Here $\mu = 1,2,..,5$  runs over the $4+1$ directions. The $\mu=1$ direction will denote the fictitious temperature direction, while the remaining directions represent Euclidean space-time. The usual staggered fermion phase factors $\eta_{\mu,x}$ obey the relations: $\eta_{1,x}^2 = T$ and $\eta_{i,x}^2 = 1$ for $i=2,3,4,5$. The parameter $T$ controls the fictitious temperature. The four fermion coupling $\tilde c$ sets the strength of the anomaly. As explained in \cite{Cecile:2007dv}, the above model has the same symmetries as $N_f=2$ QCD.

We have developed an efficient cluster algorithm to solve this model and have studied it in the $\epsilon$-regime \cite{Cecile:2007dv} and the $p$-regime \cite{Cecile:2008kp}. In our work we fix $L_t=2$ and $\tilde{c} = 0.3$. For these parameters the temperature $T$ can be tuned so that the model is either in a spontaneously broken phase for $T<T_c$ or in the symmetric phase for $T>T_c$, where $T_c=1.73779(4)$ \cite{Cecile:2007dv}. Since the phase transition is second order, close to $T_c$ the pion decay constant in the chiral limit $F$ is small in lattice units. This reduces the lattice artifacts in our model. Further, tuning $T$ close to $T_c$ in the low temperature phase also guarantees the existence of a light $\sigma$-resonance. For this reason, we choose to fix $T=1.7$ in this work.

We focus on three observables: The vector current susceptibility $Y_v$, the chiral current susceptibility $Y_c$, and the chiral condensate susceptibility $\chi_\sigma$. These are defined as
\begin{equation}
\label{yv}
Y_{v,c} = \frac{1}{d L^d}\bigg\langle 
\sum_{\mu=1}^d\bigg(\sum_x J_{\mu}^{v,c}(x)\bigg)^2\bigg\rangle,\ \ \ 
\chi_\sigma = \frac{1}{L^d} \sum_{x,y} 
\langle \overline{\psi}_x\psi_x \ \overline{\psi}_y\psi_y\rangle 
\end{equation}
where $J_\mu^{v}(x)$ and $J_\mu^c(x)$ denote one of the components of the vector and the chiral current respectively. For a detailed discussion of our algorithm and observables, we refer the reader to \cite{Cecile:2007dv}.

\section{Results}

\begin{figure}[t]
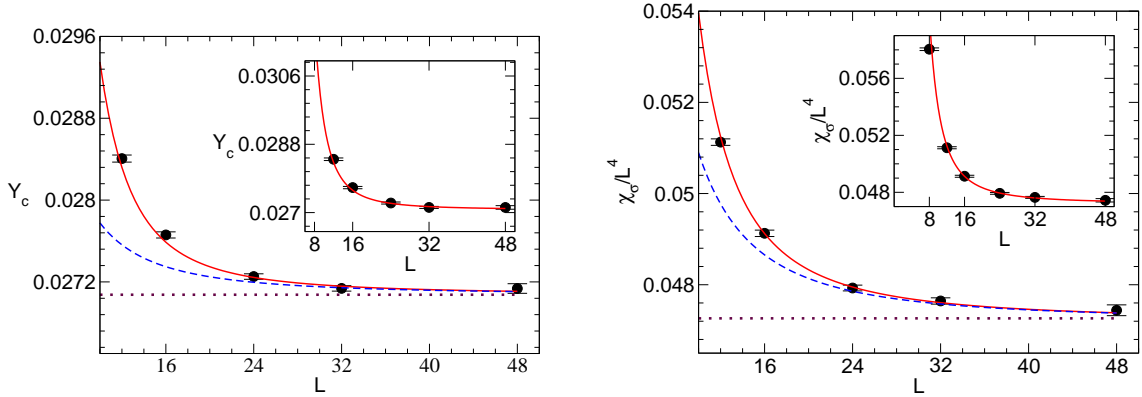

\begin{center}
\hbox{
\includegraphics[width=0.47\textwidth]{yc.eps}
\hskip0.3in
\includegraphics[width=0.47\textwidth]{chis.eps}
}
\end{center}
\caption{\label{epsi} Finite size scaling of observables in the $\epsilon$-regime. The left figure shows the current susceptibility and the right figure shows the chiral condensate susceptibility both at $m=0$. The insets show a larger range of data. The solid line is the fit to the eq.(3.1) while the dashed line is obtained when we set $a=b=0$.}
\end{figure}

We first set $m=0$ and study the finite size scaling of our observables. This is in the $\epsilon$-regime of chiral perturbation theory. At 1-loop the theory predicts that \cite{Hasenfratz:1989pk,Hansen:1990yg,Colangelo:2003hf,Colangelo:2006mp}
\begin{equation}
Y_c = Y_v = \frac{F^2}{2}\Bigg( 1 + \frac{0.14046}{(FL)^2} + \frac{a}{(FL)^4...}\Bigg), \ \ \chi_\sigma = \frac{\Sigma^2L^4}{4}\Bigg( 1 + \frac{0.42138}{(FL)^2} + \frac{b}{(FL)^4...}\Bigg)
\end{equation}
Our data and fits are shown in figure \ref{epsi}. Using the fits we extract the low energy constants $F$ and $\Sigma$ and find that $F=0.2327(1), a=1.91(9)$ with a $\chi^2/DOF=1.2$ and $\Sigma=0.4346(2), b=1.72(11)$ with a $\chi^2/DOF=0.2$. 

Next we vary the quark mass in the interval $0.0002 \leq m \leq 0.01$ for lattices in the range $12 \leq L \leq 32$ and thus explore the $p$-regime of the chiral expansion. Here the 1-loop predictions for $Y_c$, $Y_v$ and $\chi_\sigma$ are given by \cite{Hansen:1990yg,Colangelo:2006mp}.
\begin{subequations}
\label{fss}
\begin{eqnarray}
Y_c &=& (F_\pi)^2 \Bigg[1 - 2 \tilde{g_1}(L M_\pi)\xi + {\cal O}(\xi^2)\Bigg]
\\ \nonumber \\ 
Y_v &=&  (F_\pi)^2 \Bigg[ - 2 L \frac{\partial \tilde{g_1}(L M_\pi)}{\partial L} \xi + {\cal O}(\xi^2)\Bigg]
\\ \nonumber \\
\chi_\sigma &=& (\langle{\overline q}q\rangle)^2 L^4
\bigg[1 - 3 \tilde{g_1}(L M_\pi) \xi + {\cal O}(\xi^2)\bigg]  
\end{eqnarray}
\end{subequations}
where $M_\pi$ is the pion mass, $F_\pi$ is the pion decay constant and $\langle{\overline q}q\rangle$ is the chiral condensate at a given quark mass $m$. The function $\tilde{g_1}$ arises due to pions constrained to be inside a periodic box and is given by
\begin{equation}
\label{gfn}
\tilde{g}_1(\lambda) = \sum_{n_1,n_2,n_3,n_4\neq 0}^{\infty}
\frac{4}{\lambda \sqrt n} K_1(\lambda \sqrt n)
\end{equation}
where $K_1$ is a Bessel function of the second kind and $n = n_1^2 + n_2^2 + n_3^2 +n_4^2$. Fitting our data with these predictions we can determine $M_\pi, F_\pi$ and $\langle{\overline q}q\rangle$ as a function of the quark mass $m$.

\begin{table}[hb]
\begin{center}
\begin{tabular}{c| c c c} \hline\hline
\em m &\em $\langle \overline q q \rangle$ &\em $F_\pi$ &\em $M_\pi$ \\
\hline
0.0002 &0.4392(2) &0.2348(1) &0.0400(2)   \\
0.0005 &0.4441(2) &0.2377(1) &0.0627(2)   \\
0.0010 &0.4528(2) &0.2423(1) &0.0878(1)   \\
0.0020 &0.4678(2) &0.2501(1) &0.1220(2)   \\
0.0035 &0.4867(2) &0.2606(1) &0.1584(2)   \\
0.0050 &0.5024(3) &0.2690(2) &0.1860(3)   \\
0.0065 &0.5170(3) &0.2764(2) &0.2083(4)   \\
0.0100 &0.5433(2) &0.2912(2) &0.2521(5)   \\
\hline\hline
\end{tabular}
\end{center}
\caption{Results for $M_\pi$, $F_\pi$ and $\langle \overline q q \rangle$ from fitting $Y_v$,$Y_c$, and $\chi_\sigma$ as a function of $L$ to the finite-size 1-loop chiral perturbation theory. \label{tab1}}
\end{table}

\begin{figure*}[t]
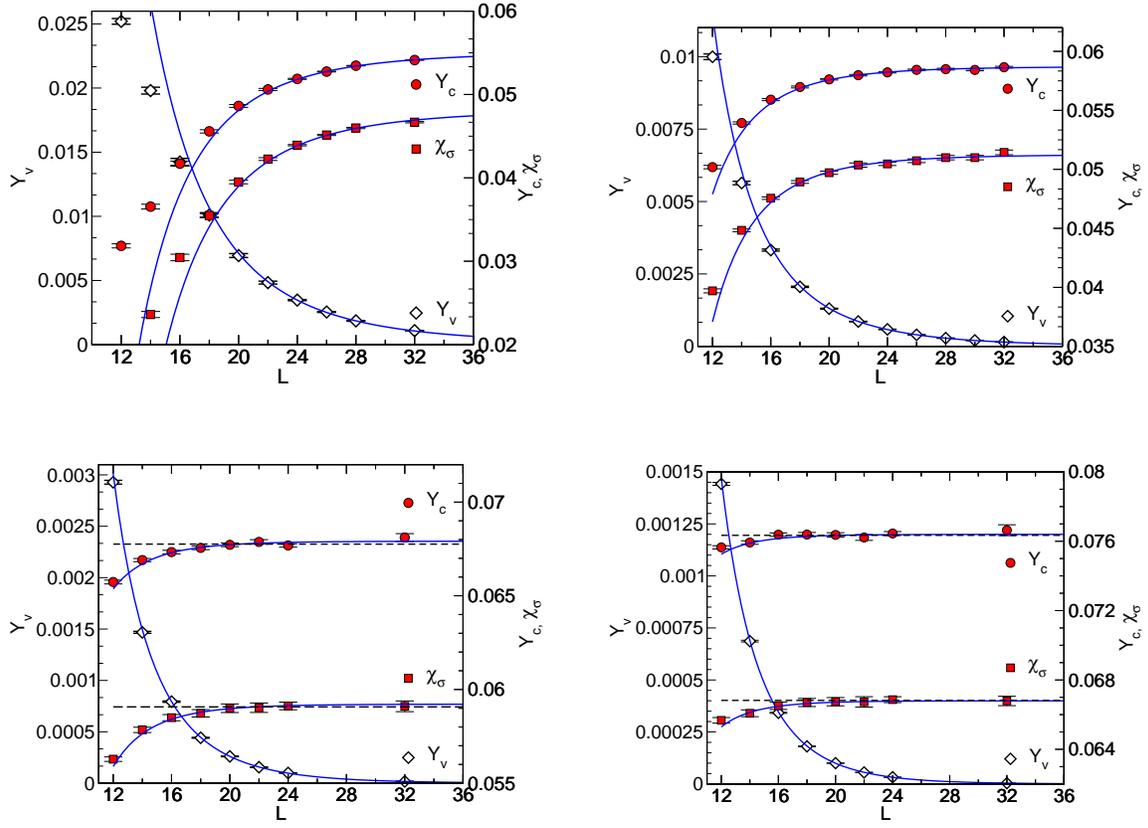

\begin{center}
\vbox{
\hbox{
\includegraphics[width=0.47\textwidth]{fig1.eps}
\hskip0.3in
\includegraphics[width=0.47\textwidth]{fig2.eps}
}
\vskip0.4in
\hbox{
\includegraphics[width=0.47\textwidth]{fig3.eps}
\hskip0.3in
\includegraphics[width=0.47\textwidth]{fig4.eps}
}
}
\end{center}
\caption{\label{preg} Finite size scaling of $Y_v$,$Y_c$ and $\chi_\sigma$ at $m=0.0002$ (top left), $m=0.001$ (top right), $m=0.0035$ (bottom left) and $m=0.0065$ (bottom right). The solid lines are fits of the data to the expected finite size scaling form from chiral perturbation theory while dashed lines are fits to a constant.}
\end{figure*}

Our data fit well to the above predictions for $0.0002 < m \leq 0.0035$. For $m > 0.0035$ the fits converge only if we exclude almost all the curvature in $Y_c$ and $\chi_\sigma$. In particular, we are not sensitive to the $\tilde{g}_1(\lambda)$ function for these two observables and the data fit well even to a constant.  On the other hand $Y_v$ continues to fit well for the entire range of data and this can be used to extract $M_\pi$ accurately as a one parameter fit. This may be a useful observation even for lattice QCD calculations. Thus, we were able to extract $F_\pi$, $M_\pi$ and $\langle {\overline q}q\rangle$ as functions of the quark mass. Some of our results are summarized in Table~\ref{tab1}. As an illustration, we also show our data at $m=0.0065,m=0.0035,m=0.001$ and $m=0.0002$ along with the fits in Fig.~\ref{preg}.

\begin{table}[b]
\begin{center}
\begin{tabular}{c  c   c   c   c | c } \hline\hline
$\Sigma$ & $F$ &\em $\cS$ & $\cF$ & $\cM$ &\em $\chi^2$
\\\hline
0.4354(3) &0.2329(2) &11.9(3) &19.3(5) &39(3) &1.1    \\
\hline\hline
\end{tabular}
\end{center}
\caption{ \label{tab2} Results from a combined fit of the data in Table 1 to Eqs.(3.6)}
\end{table}

The quark mass dependence of $F_\pi$, $\langle \overline{q}q\rangle$ and $M_\pi$ have also been computed up to 1-loop in \cite{Hansen:1990yg,Hasenfratz:1989pk}:
\begin{equation}
\label{infvol}
F_\pi = F\bigg[1 -\xi'(\log\xi' - 2\cF)\bigg];\ 
\langle{\overline q}q\rangle = 
\Sigma\bigg[1 - \frac{3}{2}\xi'(\log\xi' - 2 \cS \bigg];\ 
M^2_\pi = M^2 
\bigg[1 +\frac{1}{2}\xi'(\log\xi' - 2 \cM)\bigg],
\end{equation}
where $\cF,\cS$ and $\cM$ are higher order low energy constants and are usually defined in the literature as $c_i = \log(\Lambda_i/4\pi F)$. We have performed a combined fit of $F_\pi$,$\langle{\overline q}q\rangle$ and $M_\pi$, quoted in Table~\ref{tab1}, in the region $0.0002 \leq m \leq 0.001$ to the above three relations. The result is tabulated in Table~\ref{tab2}. The values of $F$ and $\Sigma$ obtained in the $p$-regime agree nicely with the those computed in the $\epsilon$-regime. 

\begin{figure*}[t]
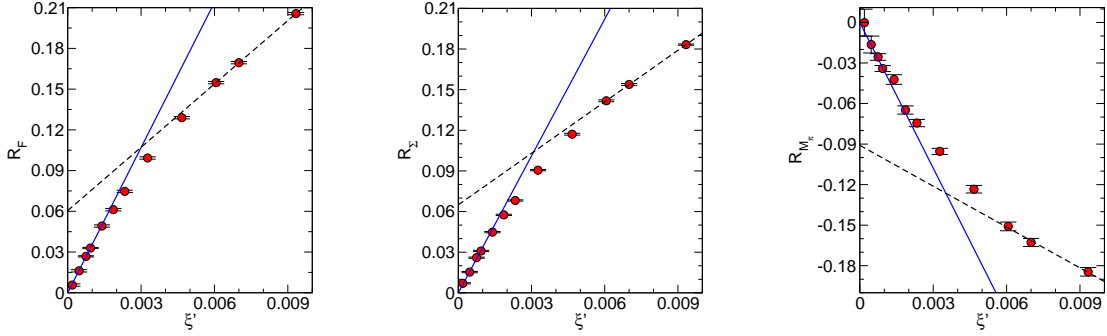

\begin{center}
\hbox{
\includegraphics[width=0.27\textwidth]{fig5.eps}
\hskip0.4in
\includegraphics[width=0.27\textwidth]{fig6.eps}
\hskip0.4in
\includegraphics[width=0.28\textwidth]{fig7.eps}
}
\end{center}
\caption{\label{fig3} 
Rescaled and subtracted quantities defined in the text which go to zero linearly in the region where 1-loop chiral perturbation theory is valid.  The solid lines are plots of the fits discussed in the text. The dashed lines show the linear region for larger values of $\xi'$. The ``knee'' is estimated roughly as the point where the two lines cross.}
\end{figure*}

In order to isolate the region where 1-loop corrections are a good description of the data we define the following rescaled and subtracted quantities: $R_{F} \equiv F_\pi/F - 1 + \xi'\log\xi'$, $R_{\Sigma} \equiv \langle{\overline q}q\rangle/\Sigma - 1 + 3\xi'\log\xi'/2$, and $R_{M} \equiv M_\pi^2/M^2 - 1 - \xi'\log(\xi')/2$ using $F=0.2329$ and $\Sigma =0.4354$ obtained from our fits. By definition, the $R$'s must be linear in $\xi'$ in the region where 1-loop results are valid. In Fig.~\ref{fig3} we plot the $R$'s as a function of $\xi'$. Assuming errors of $1\%$ or less can be tolerated, Fig.~\ref{fig3} shows that 1-loop chiral perturbation theory describes the data for $\xi' \lesssim 0.002$. Interestingly, there is also an approximately linear region for $\xi' \gtrsim 0.006$ but with a completely different slope. This is shown as the dashed line in Fig.~\ref{fig3}. This behavior suggests that chiral perturbation theory begins to break down roughly around $\xi' \approx 0.0035$, which is the location of the ``knee'' that separates the low $\xi'$ and high $\xi'$ regions. We will argue below that the $\sigma$-resonance is responsible for this break down.
 
\section{Discussion and Conclusions}

It has been argued in the context of the $O(4)$ linear sigma model, that the physics in the sigma channel is directly related to the coefficients $\cF, \cS$ and $\cM$. Perturbative calculations show that \cite{Gockeler:1992zj,Gockeler:1991sj,Hasenfratz:1990fu}:
\begin{equation}
\label{sigma}
\cS = \log(M_R/4\pi F) - \frac{7}{6} + \frac{8\pi^2}{3 g_R},\ \ \ 
\cM = \log(M_R/4\pi F) - \frac{7}{3} + \frac{8\pi^2}{g_R} 
\end{equation}
where $M^2_\sigma = M^2_R [1 + g_R(3\pi\sqrt 3 -13)/(16\pi^2)]$. Here $M_\sigma$ is that mass of the $\sigma$ particle and $g_R$ is the corresponding renormalized coupling, $g_R = M_R^2/2 F^2$. Since our model is close to the $O(4)$ phase transition the linear sigma model should be a reliable description of the physics in the sigma channel. Indeed, using $\cS = 12$ we find that $M_\sigma/F \sim 2$ while using $\cM = 39$ we again find that $M_\sigma/F \sim 2$. The fact that these two agree with each other is a confirmation of our belief. Assuming $M_\sigma/F \sim 2$ and setting the scale of our lattice with $F = 90$MeV we estimate $M_\sigma \sim 180$MeV in our model. At $\xi' \sim 0.0035$ we find that $M_\pi \sim 60$MeV. Hence, we conclude that when $M_\pi > M_\sigma/3$ chiral perturbation theory begins to break down and the physics is better described by the linear sigma model.

There are many important differences between our model and QCD. However the most important difference is that our model contains a light and perhaps a narrow $\sigma$-resonance while in QCD the $\sigma$-resonance is expected to be heavier and broader. This is the reason why our low energy constants turned out to be much larger than QCD and the convergence of chiral perturbation theory was affected. Despite the differences, it is indeed encouraging that our model supports the most important expectations of chiral perturbation theory namely, chiral perturbation theory is applicable in a region of small quark masses and that the properties of resonances play an important role in determining this region. In particular, we have found evidence that the properties of $\sigma$-resonance are encoded in the low energy constants that control the chiral logarithms and hence can play an important role in determining the region where 1-loop chiral perturbation theory is valid. The resonance properties of course change with the quark masses. Thus, it is safe to assume that chiral perturbation theory can only become reliable in the region of the quark mass where the properties of the $\sigma$-resonance (and other resonances) varies little. A rough estimate based on \cite{Hanhart:2008mx} suggests that $M_\pi \lesssim 250$MeV may be necessary. Thus, it should not be very surprising that 1-loop chiral perturbation theory may be applicable at a few percent accuracy only at realistic pion masses.

\end{document}